\documentclass[twocolumn,aps,pra,nofootinbib,groupedaddress,amsmath,amssymb,longbibliography]{revtex4-1}
\usepackage{graphicx}
\usepackage{dcolumn}
\usepackage{bm}
\usepackage{color}
\usepackage{multirow}
\usepackage[mathscr]{euscript}
\usepackage{subfigure}
\usepackage[colorlinks=true,allcolors=blue]{hyperref}
\usepackage{braket}
\usepackage{lipsum}
\newcommand{\RNum}[1]{\uppercase\expandafter{\romannumeral #1\relax}}

\newcommand{\BE}{\begin{eqnarray}}
\newcommand{\EE}{\end{eqnarray}}
\newcommand{\dagg}{^{\dagger}}
\newcommand{\1}{&=&}
\newcommand{\2}{&+&}

\newcommand{\ee}[1]{\mathrm{e}^{#1}}
\newcommand{\nn}{\nonumber}
\newcommand{\ii}{i}
\newcommand*\dif{\mathop{}\!\mathrm{d}}

\newcommand{\figtop}[4]{
 \begin{figure}[!t]
 \begin{center}
 \scalebox{#3}{\includegraphics{#1}}
 \vspace{-0.1in}
 \caption{\label{#4}#2}
 \end{center}
 \end{figure}
}

\begin{document}
\title{Nonlinear dynamics of Rydberg-dressed Bose-Einstein condensates in a triple-well potential}


\author{Gary McCormack$^{1}$, Rejish Nath$^{2}$ and Weibin Li$^{1}$}

\affiliation{$^{1}$School of Physics and Astronomy, and Centre for the Mathematics and Theoretical Physics of Quantum Non-Equilibrium Systems, University of Nottingham, NG7 2RD, United Kingdom\\ $^2$Indian Institute of Science Education and Research, Pune, 411008, India}%

\begin{abstract}
We study nonlinear dynamics of Rydberg-dressed Bose-Einstein condensates (BECs) trapped in a triple-well potential in the semiclassical limit. The Rydberg-dressed BECs experience a long-range soft-core interaction, giving rise to strong nearest and next-nearest neighbor interactions in the triple-well system. Using mean-field Gross-Pitaevskii (GP) equations,  we show that lower branches of the eigenspectra exhibit loops and level-crossings when the soft-core interaction is strong. The direct level-crossings eliminate the possibility of adiabatic Landau-Zener transitions when tilting of the triple-well potential.
We demonstrate that the long-range interaction allows for self-trapping in one, two, or three wells, in a far more controllable manor than BECs with short-range or dipolar interactions. Exact quantum simulations of the three-well Bose-Hubbard model indicate that self-trapping and nonadiabatic transition can be observed with less than a dozen bosons. Our study is relevant to current research into collective excitation and nonlinear dynamics of Rydberg-dressed atoms.
\end{abstract}

\maketitle

\section{Introduction}\label{sec:Introduction}
The understanding of the dynamics of interacting Bose-Einstein condensates (BECs) has been a lucrative field of research in the past three decades ~\cite{Polkovnikov2002a,Anderson1995,Viscondi2011,Wang2006,PitaevskiiLev2016BCaS,Pethick2008a}. With modern experimental techniques that allow for controlling properties of ultracold atomic gases, such as atom-atom interactions~\cite{Smith2013}, trapping potentials and spatial dimensions~\cite{Fallani2005,Smerzi1997,Zenesini2010}, along with long coherence times~\cite{Bell2016}, stationary and dynamical properties of atomic BECs have been explored in great detail~\cite{PitaevskiiLev2016BCaS}. The dynamics of a trapped atomic BEC is typically described by the mean-field, Gross-Pitaevskii (GP) equation~\cite{Smith2013}, with which many interesting properties and novel dynamics have been revealed~\cite{Muryshev2002,Brazhnyi2004,Fritsch2020,Li2008,Rogel-Salazar2013,Boccato2018,McKay2015,Witthaut2011,Li2018b}. In optical lattices~\cite{Jaksch2005}, bosons can undergo the well-known superfluid-Mott insulator transition~\cite{Bloch2005}. It has been proposed that BECs~\cite{Byrnes2012,Byrnes2015,Dadras2018} and atoms trapped in optical lattices~\cite{brennen_quantum_1999} can be used for carrying out quantum computation.

\begin{figure}[t!]
\centering
\includegraphics[width=0.9\linewidth]{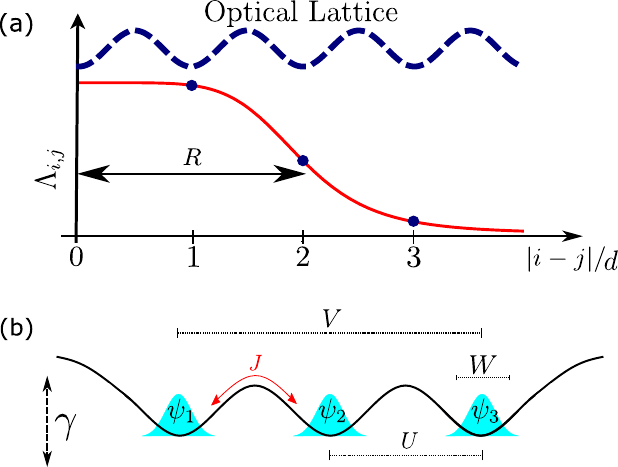}
\caption{(Color online) \textbf{Long-range soft-core interaction and finite lattice trapping potential}. (a) The soft-core interaction potential (red) extends across several lattice sites. The soft-core radius can be larger than the lattice constant $d$ of the optical lattice potential (dashed). When the atomic separation is larger than $R$, the interaction decreases rapidly. (b) Schematic of a triple-well lattice potential. The potential height (zero-point energy) $\gamma$ of each trap may be adjusted dynamically. 
Atoms may interact via onsite ($W$), nearest ($U$) or next-nearest ($V$) neighbor interactions, while the tunneling $J$ is restricted to nearest-neighbor sites, forming a chain setup. }
\label{fig:triplewell}
\end{figure}

Substantial work has also been carried out in finite-sized double-well and triple-well potentials. In the quantum regime, the dynamics of atoms in double well potentials are affected by the onsite (short-range) interactions, causing wave-packet collapse and revivals~\cite{Milburn1997}. In the semiclassical regime, strong onsite interactions introduce interesting nonlinear effects. One striking feature is that the eigenspectrum of the nonlinear system develops a loop structure due to strong onsite interactions~\cite{Liu2002a,Liu2003a}. The loop causes the breakdown of the adiabatic theorem and self-trapping dynamics, which has been examined experimentally~\cite{Albiez2005,Zibold2010}.
In triple-well configurations, static and dynamical properties depend on boundary conditions and spatial profiles of two-body interactions. With closed boundaries and bare onsite interactions~\cite{Graefe2006a,Liu2007,Liu2018}, multiple loop structures are found in the spectra of the coupled nonlinear system. These complicated spectra lead to turbulent phase spaces~\cite{Dey2018,Dey2019,Burkle2019} and produce oscillatory dynamics beyond the typical Josephson or self-trapping behavior~\cite{Liu2007,Graefe2006a,Wang2017}. Population transfer between energy levels has been found via Landau-Zener tunneling~\cite{Graefe2006a,Dey2018,Dey2019}.
Ring-shaped triple-well setups (i.e. periodic boundary conditions) have also been examined, in which eigenenergies intersect even for the noninteracting case~\cite{Guo2014}.

When long-range dipolar interactions are present, the dynamics of BECs are changed dramatically in triple-well potentials. The ground state shows exotic behavior, such as mesoscopic quantum superpositions \cite{Lahaye2010} and macroscopic first-order coherence between the outer sites \cite{Xiong2013}.  Recently, quantum population and entanglement dynamics of dipolar BECs in triple wells have also been examined \cite{Tonel2020}. It was shown that the nonlocal interactions allow both coherent and non-coherent oscillations between the sites, with very little dynamics occurring in the middle well. However, dipolar interactions decay rapidly with distance $r$ as $r^{-3}$, which leads to weak nearest-neighbor and much weaker next-nearest-neighbor interactions in a triple-well potential.

Long-range interactions can be realized alternatively by dressing ground state atoms to electronically high-lying Rydberg states, leading to soft-core shaped long-range interaction potentials~\cite{Bouchoule2002,Henkel2010a,Honer2010,Pupillo2010,Johnson2010,Li2012}. The soft-core interaction is nearly a constant within a radius $R$. For typical parameters, the soft-core radius is a few micrometers~\cite{Henkel2010a} after which the dressed interaction decreases as $r^{-6}$ as depicted in Fig. \ref{fig:triplewell}(a).
This interaction has motivated a number theoretical studies on the static and dynamical properties of Rydberg-dressed atoms confined in traps~\cite{Henkel2012,Maucher2011c,Balewski2014,Cinti2014,Xiong2014,Hsueh2016,McCormack2020} and optical lattices~\cite{Lauer2012,Lan2015,Li2015,Angelone2016,Chougale2016,Li2018a,Zhou2020}. Additionally, recent experiments have successfully demonstrated Rydberg-dressing in optical tweezers~\cite{Jau2016b}, optical lattices~\cite{Zeiher2016,Zeiher2017,guardado_sanchez_quench_2020}, and traps~\cite{Li2018b,Borish2020}.

In this work, we study BECs interacting with long-range soft-core interactions trapped in a triple-well potential [see Fig.~\ref{fig:triplewell}(b)]. A key feature is that the long-range Rydberg-dressed interaction allows us to explore dynamics in a regime where nearest-neighbor and next-nearest-neighbor interactions are strong, due to the large soft-core radius. When the traps are tilted, the system undergoes nonadiabatic Landau-Zener transitions due to complicated loops and level-crossings on the lower branches of the eigenspectra.
This results in dynamical instability and hence leads to the breakdown of the adiabatic theorem.  This is in stark contrast to systems with short-range interactions, where tunneling from the ground state is not prevented from adiabatic population dynamics, as the level-crossings emerge in the higher energy branches.
By tuning the profile of Rydberg-dressed interactions, we can also control self-trapping of BECs~\cite{Smerzi1997,Raghavan1999} to a high degree of accuracy, which is typically difficult if considering only onsite interactions. We propose that the nonlocal interactions allow for precise manipulation of the final states, such that we can control whether the trapping is localized in a one, two, or even all three wells simultaneously. We also carry out simulations of the quantum dynamics which takes into account the inter-well correlations. The comparison with the mean-field results show that the transporting dynamics can be found in mesoscopic systems with tens of atoms.

The paper is organized as follows. In Sec. \ref{sec:Hamiltonian} the Hamiltonian of the system is introduced. The corresponding mean-field approximation is presented and the resulting equations of motion are given. We examine the eigenspectrum and discuss new features in our system.
In Sec \ref{sec:Results}, we explore nonadiabatic transitions for both weak and strong nonlinear interactions. The Landau-Zener transition probability is also examined.
By analyzing the Poincar\'e sections for different energy values, we show that the system can move towards highly chaotic regions when the nonlinear interactions are strong.
We then examine self-trapping of bosons in different sites. The dynamics depends on initial conditions and long-range interactions. We moreover compare the mean-field results to quantum dynamics.  We conclude our work in Sec. \ref{sec:Conclusion}.

\section{Model and Method}\label{sec:Hamiltonian}
\subsection{Bose-Hubbard and mean-field Hamiltonian}
We consider $N$ bosonic atoms in a one-dimensional trap array, whose dynamics is governed by an extended Bose-Hubard Hamiltonian ($\hbar=1$)
\BE
\label{Ham:bh}
\hat{H} &=& \sum_j^{L}\Gamma_j\hat{n}_j-J\sum_{\langle i,j\rangle}^{L}\hat{a}_i\dagg\hat{a}_j\nn\\
&& +\frac{g}{2}\sum_{j}^{L}\hat{n}_j(\hat{n}_j - 1)+\frac{1}{2}\sum_{i, j}^{L}\Lambda_{i,j}\hat{n}_i\hat{n}_j,
\EE
where $J$ and $L$ are the hopping rate of atoms between nearest-neighbor traps and total number of traps, respectively.
In this work we imagine a triple-well chain setup ( i.e $L=3$ with closed boundary conditions), where we restrict hopping to nearest neighbors only, denoted by $\langle ...\rangle$.
The bosonic annihilation (creation) operator at site $j$ is given by $\hat{a}_j (\hat{a}_j^{\dagger})$.
$\Gamma_j$ and $\hat{n}_j=\hat{a}_j\dagg\hat{a}_j$ is the local tilting potential and number operator, respectively. The parameter $g=4\pi a_s/m$ characterizes the onsite (s-wave) interaction~\cite{PitaevskiiLev2016BCaS,Pethick2008a}, whose value can be controlled through Feshbach resonances~\cite{Makotyn2014a}; where $a_s$ and $m$ are the scattering length and mass, respectively. $\Lambda_{i,j}=C_6/[|i-j|^{6}d^{6} + R^{6}] $ is the soft-core interaction between site $i$ and $j$ with $d$ being the lattice constant, and $C_6$ being the van der Waals coefficient~\cite{Henkel2010a}.

In the limit of $N\gg 1$, we employ the mean-field approximation to replace the bosonic operator with a classical field $\psi_{j}$, i.e. $\hat{a}_j\approx\psi_j$, $\hat{a}_j^{\dagger}\approx\psi_j^{*}$ and $\sum_{j}|\psi_j|^2=N$~\cite{Smith2013}. This yields the mean-field Hamiltonian
\BE
\label{Ham:mf}
\tilde{H}&=&\sum_j^3\Gamma_j|\psi_{j}|^2-J\sum^3_{\langle i,j\rangle}\psi_{i}^*\psi_{j}\nn \\
&&+\frac{g}{2}\sum^3_j |\psi_j|^2(|\psi_j|^2-1) + \frac{1}{2}\sum^3_{i, j}\Lambda_{i,j}|\psi_{i}|^2|\psi_{j}|^2. 
\EE
The dynamics of the classical field $\psi_i$ is derived via the canonical equation $id\psi_{j}/dt=\partial \tilde{H}/\partial \psi_{j}^*$. For convenience, we define the normalized field $c_{j}=\psi_{j}/\sqrt{N}$ with the normalization condition $\sum_{j}|c_{j}|^{2}=1$. For the triple-well system, we obtain the following coupled nonlinear GP equations
\begin{subequations}
\BE
i \dot{c}_1 \1  \left(W |c_1|^2 + U|c_2|^2+V|c_3|^2 \right)c_1 +\gamma  c_1 - Jc_2 ,\label{eqm1}\\
i \dot{c}_2 \1 \left[W|c_2|^2 + U(|c_1|^2+|c_3|^2)\right] c_2 -J(c_1+c_3) ,\label{eqm2}\\
i \dot{c}_3 \1 \left(W|c_3|^3 + U|c_2|^2+V|c_1|^2 \right)c_3 -\gamma c_3 - Jc_2   ,\label{eqm3}
\EE
\end{subequations}
where we have  defined $W=N(\Lambda_{11} + g)$, $U=N\Lambda_{1,2}$ and $V=N\Lambda_{1,3}$ to be the onsite, nearest-neighbor and next-nearest-neighbor mean-field interactions, respectively. The short-range interaction $W$ takes into account of contributions from the s-wave and onsite soft-core interaction. The local potential $\Gamma_j$ is antisymmetric, given by $\Gamma_j=-(j-2)\gamma$, i.e. $\Gamma_1= \gamma$, $\Gamma_2=0$, and $\Gamma_3=-\gamma$. Here $\gamma$ is a bias field to create a potential height difference between neighboring traps. In Sec. \ref{subsec:LZ}, the potential wells are linearly biased through $\gamma=\alpha t$, with $\alpha$ being the sweep rate. In Sec. \ref{subsec:ST}, we will consider a fixed $\gamma$.
To be convenient, we will scale time and energy with respect to $1/J$ and $J$ in the following unless stated explicitly.

\subsection{Adiabatic eigenspectra of the GP equation}
\begin{figure}
\centering
\includegraphics[width=0.97\linewidth]{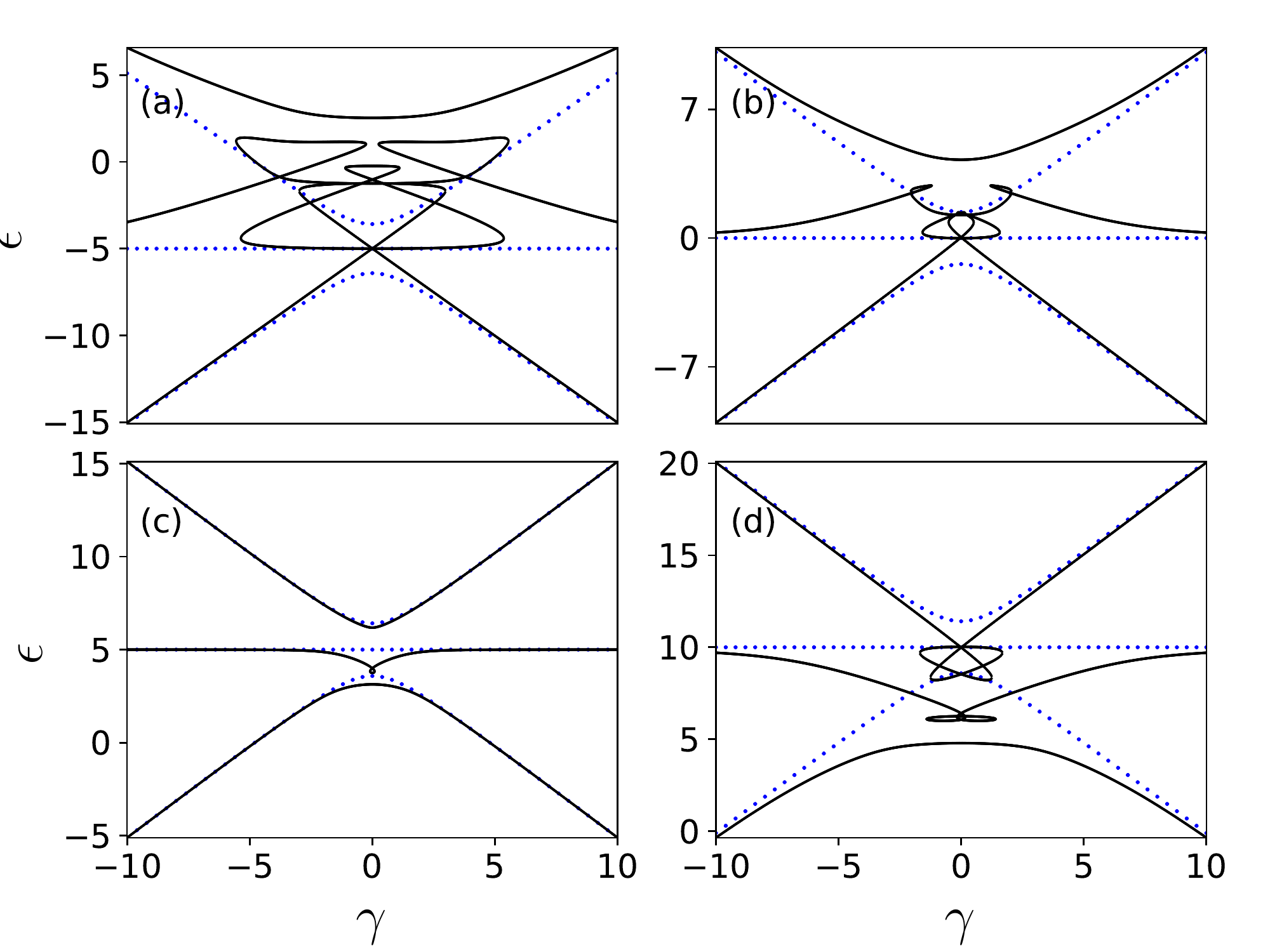}
\caption{(color online) \textbf{Adiabatic eigenspectra with different short-range interactions}. We show the adiabatic eigenspectra as a function of $\gamma$ for (a) $W=-5$, (b) $W=0$, (c) $W=5$, and (d) $W=10$ while fixing $U=2V=5$. When the short-range interaction is attractive or vanishing, loops and direct level-crossings are found in the lower branches. These structures disappear when $W=U$. When the short-range interaction dominates, the structures are found in the upper branches of the levels. The linear case ($W=U=V=0$) is shown for reference in each panel (blue dotted). To compare with the nonlinear spectra, the linear spectra are shifted perpendicularly by $W$.}\label{fig:short}
\end{figure}
\figtop{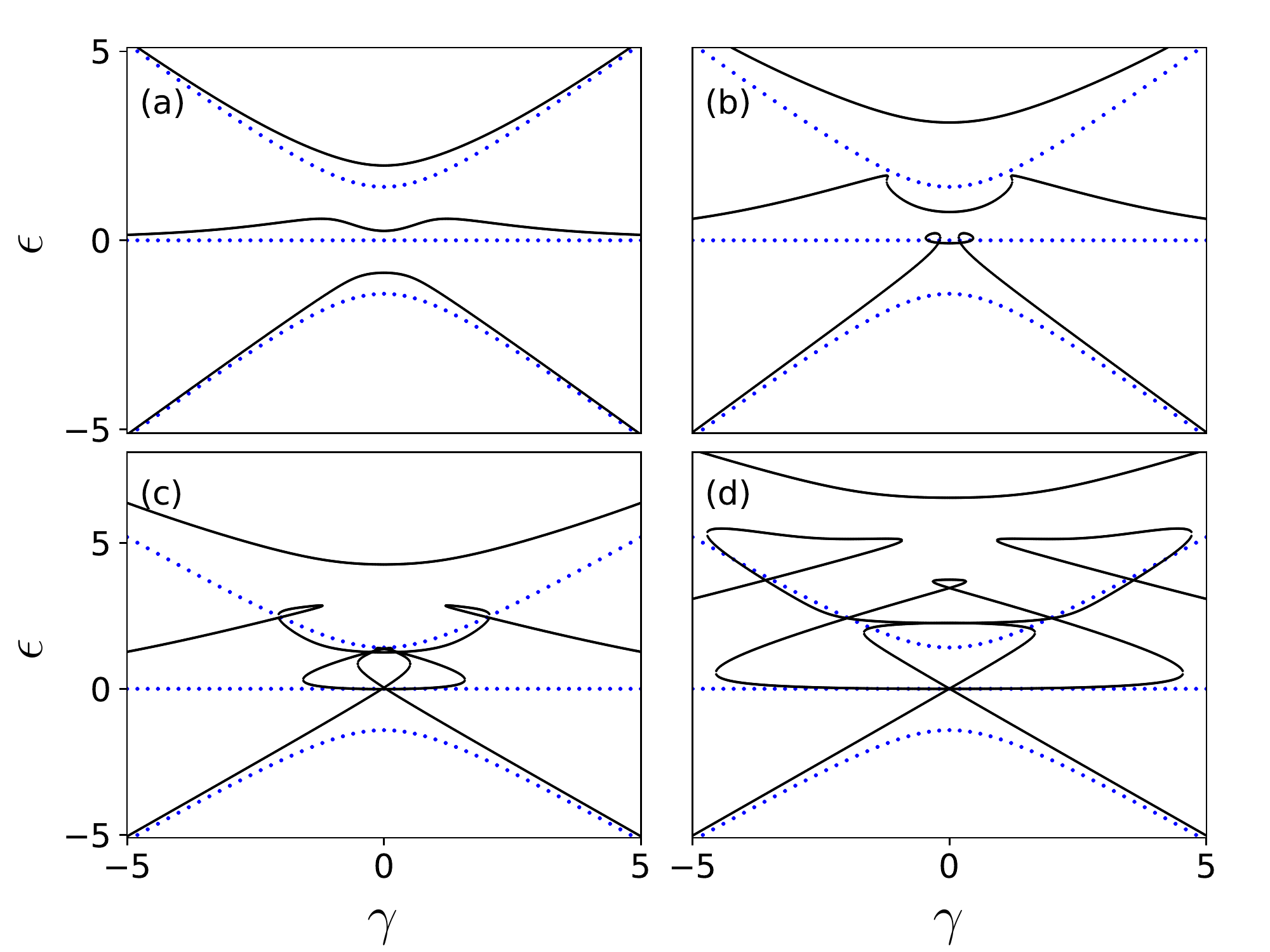}{(Color online) \textbf{Adiabatic eigenspectra with different long-range interactions}. By turning off the short-range interactions ($W=0$), the  adiabatic eigenspectra are shown as a function of $\gamma$ for $U=2V$ with (a) $U=1$, (b) $U=3$, (c) $U=5$, and (d) $U=9$.  When the long-range interactions are strong, loops and direct level-crossings emerge on the lower branches, as shown in panels (b)-(d). The linear case ($W=U=V=0$) is shown for reference in each panel (blue dotted).}{0.42075}{fig:eig}
When analyzing the adiabatic spectra of the system through Eqs. (\ref{eqm1}-\ref{eqm3}), the presence of the nonlinearity implies that standard methods (i.e. diagonalization of the Hamiltonian) are not valid. We adapt the method applied to treat nonlinear equations corresponding to interacting BECs in double-well potentials~\cite{Liu2002a}. The basic idea is to convert the nonlinear equations into a high-order $L^2$ polynomial equation of eigenvalue $\epsilon$ additionally applying the normalization condition~\cite{Wu2000}. For $L=3$, it becomes difficult to solve the resulting polynomial equation even numerically. As such, we employ a shooting method that is similar to obtaining bound states of the Schr\"odinger equation. A trial energy $\epsilon_t$ is fed into the nonlinear GP equations, allowing us to calculate eigenvectors $[c_1,c_2,c_3]$ and eigenenergy $\epsilon_n$. An eigenstate is identified if the calculated and trial energy are equal, i.e. $\epsilon_t=\epsilon_n$. This is carried out for a fine grid of trial energies to obtain all eigenenergies.

We first investigate the interplay between short-range ($W$) and long-range ($U$ and $V$) interactions. When both the short and long-range interaction are perturbative with respect to $J$, the eigenspectra are separated and display avoid level-crossings even when $\gamma\sim0$ [see demonstration in Appendix~\ref{appendix:SR_vs_LR}]. To highlight the roles played by the nonlinear interaction, we calculate the eigenspectra of the GP equation by varying $W$ while fixing $U=2V=5$, shown in Fig.~\ref{fig:short}. When the tilting is large, i.e. $|\gamma|> |W|, \text{ }U, \text{ and }V$ the eigenspectra approaches the linear spectra. When $W=-10$  (i.e. attractive onsite interactions), we find direct level-crossings between the lowest three branches at $\gamma=0$. Slightly away from  $\gamma=0$, large loop structures are found, as shown in Fig.~\ref{fig:short}(a). When $W=0$ similar structures are found, where the sizes of the loops shrink [Fig.~\ref{fig:short}(b)].
\figtop{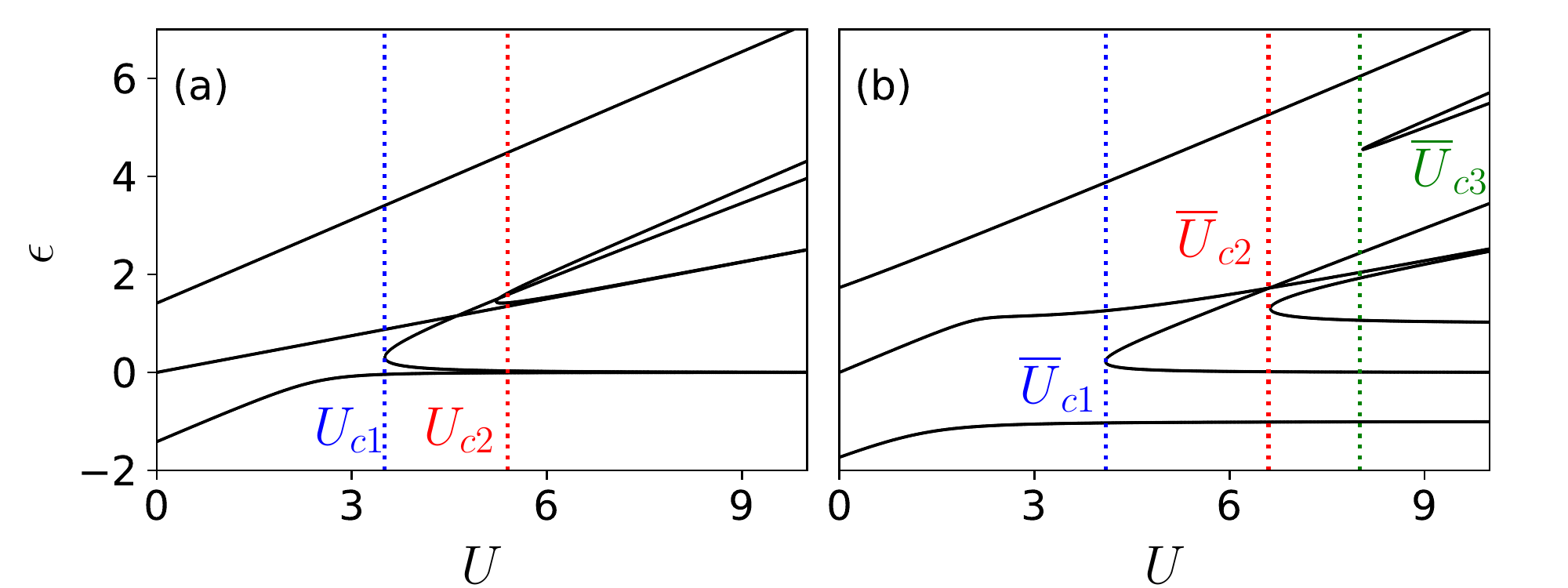}{(Color online) \textbf{Eigenspectra vs the long-range interactions}. Increasing the interaction strength leads to the creation of new energy levels in (a) symmetrical traps ($\gamma=0$) and (b) titled traps ($\gamma=1$). Location of the critical interaction strengths can be extrapolated from when a new level is created. In (a) the critical values are $U_{c1}\approx3.5$ and $U_{c2}\approx5.4$. In (b) the critical values are $\overline{U}_{c1}\approx4.1$, $\overline{U}_{c2}\approx6.6$, and $\overline{U}_{c3}\approx8$.}{0.44}{fig:eig_vsU}
 When $W=U=2V=5$, the loops disappear and the spectra is similar to the linear spectra.
This set of parameters largely gives a global energy shift. Due to the normalization condition, only Eq.~(\ref{eqm1}) and Eq.~(\ref{eqm3}) have a single nonlinear term proportional to $V$ while Eq.~(\ref{eqm2}) has no nonlinear interaction term anymore. When $W=10$ [Fig.~\ref{fig:short}(d)], the loop and level crossings re-appear in the higher energy branches. In this case the spectra are dominated by the short-range interaction.  This structure is similar to previous studies in systems with bare onsite interactions~\cite{Liu2018,Liu2007,Liu2018} where the loops and level-crossings form star-like structures.

In the remainder of this work, we will focus on a regime where only \textit{long-range interactions} are present (i.e. $W=0$). For soft-core interactions where $R\gg1$, the nearest and next-nearest neighbor interactions are the same as $U=V$. The resulting spectra can be obtained from BECs with bare short-range interactions [see details in Appendix \ref{appendix:SR_vs_LR}]. This can be understood that the particle conservation maps the long-range interaction to an attractive short-range interaction. To violate this symmetry, we will focus on a condition $U=2V$, which will be used for the remainder of this article. This restriction can move the loop and level-crossings to the central region, leading to interesting dynamics.
For weak long-range interactions, the eigenenergies are slightly modified from the linear counterpart [Fig.~\ref{fig:eig}(a)]. When the long-range nonlinear interaction is strong (i.e. $U,\,V\gg 1$), the energy levels are pushed upwards by increasing $U$ and $V$, as can be seen in Fig. \ref{fig:eig}(b). The spectrum develops a loop structure on the lowest level when $U=2V=3$. By further increasing the long-range interactions [see Figs. \ref{fig:eig}(c) and (d)], the loops become larger and more complicated level-crossings  emerge in higher energy states.

In Fig. \ref{fig:eig_vsU}(a) we plot the energy levels as a function of $U$ with $\gamma=0$.
At a critical interaction strength $U_{c1}\approx3.5$, a new branch of levels emerges. Further increasing $U$ to $U_{c2}\approx5.4$, a second branch appears at higher energies. Note that explicit values of $U_{c1}$ and $U_{c2}$ depend on $\gamma$. Fig. \ref{fig:eig_vsU}(b) shows another example for a tilted trap with $\gamma=1$. The levels are more separated in the low energy region. Here the two critical values are $\overline U_{c1}\approx4.1$ and $\overline U_{c2}\approx6.6$. Furthermore we see the emergence of a third energy level at $\overline U_{c3}\approx8$. In the following sections, we will show that the critical values of the long-range interaction strongly relate to dynamical behaviors of the system.

\section{Results and Discussion}\label{sec:Results}
\subsection{Landau-Zener and nonadiabatic transitions}\label{subsec:LZ}
In this section we study the dynamics of the long-range interacting BEC when the traps are tilted at different rates $\alpha$. Without nonlinear interactions ($U=V=0$), the level spacing is determined by the tunneling rate $J$. In the diabatic regime where $\sqrt{\alpha}$ is large and comparable to the typical level spacing $\Delta E$ [see dashed curve in Fig.~\ref{fig:eig}], the system does not have enough time to respond to the change of the tilting. Starting from the ground state, higher energy levels will be excited. In the opposite, adiabatic limit where $\sqrt{\alpha}$ is small, the adiabatic theorem states that the system will remain in an instantaneous eigenstate under variation of $\alpha$~\cite{Kato1950,Avron1999}. This has been studied extensively with two-state (well) systems, where the transition probability from $t\to -\infty$ to $t\to \infty$ is given analytically by,
\BE
P(\alpha)=\exp{\left(\frac{-2 \pi J^2}{\alpha}\right)}\label{eq:LZ}.
\EE
The resulting Landau-Zener dynamics \cite{Landau1932,Zener1932} has produced a vast field of research and is still lucrative in terms of its modern day applications.
\figtop{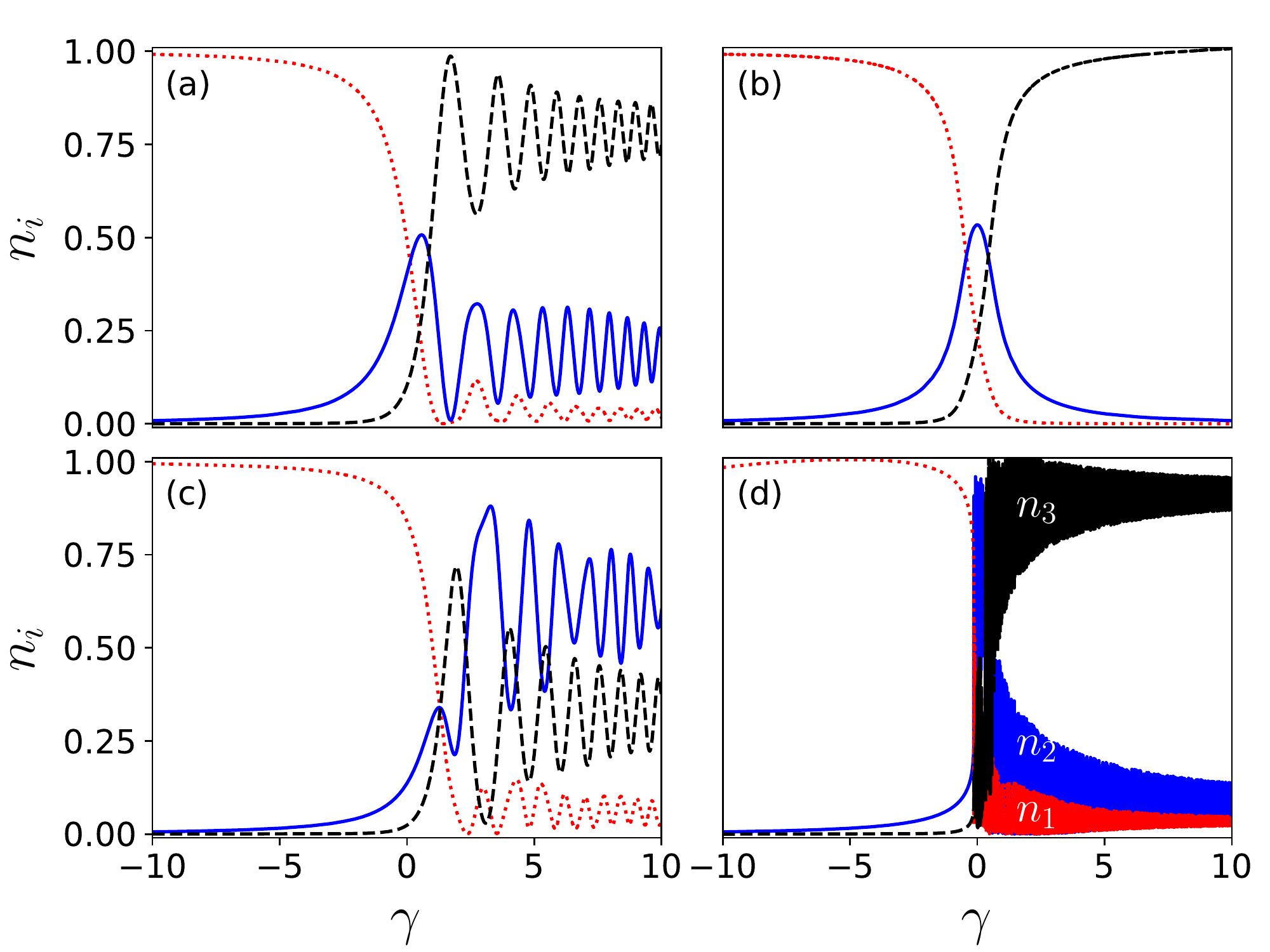}{(color online) \textbf{Landau-Zener transitions with weak and strong interactions}.  The bias potential is tilted with a fast rate $\alpha=1$ in (a) and (c), and a much slower rate $\alpha=0.001$ in (b) and (d). The interaction is $U=2V=1$ in the upper panels and $U=2V=3$ in the lower panels. For rapid tilting, higher energy modes are excited such that oscillatory dynamics are observed in (a) and (c). For slow tilting, the dynamics depends strongly on the nonlinear interactions. When the interaction is strong, the adiabatic condition is broken [see Fig.~\ref{fig:eig_vsU}(b)]. The densities $n_1$, $n_2$, and $n_3$ are given by the dotted red, solid blue, and dashed black lines. }{0.40}{fig:Ad_dynamU1}

In Fig.~\ref{fig:Ad_dynamU1}, population evolution of the BECs in the three wells is shown in the presence of weak interactions. Initially all atoms are in the left well [$n_1(0)=1$]. When the trap is tilted rapidly at rate $\alpha=1$ [Fig.~\ref{fig:Ad_dynamU1}(a)], the population undergoes fast oscillations when the tilting is reversed, i.e. $\gamma>0$. This case corresponds to the diabatic regime where the lowest energy gap is $\Delta E=0.6$, comparable to $\sqrt{\alpha}=1$. The level spacing $\Delta E$ now depends on the nonlinear interaction strengths, in addition to the hopping. In Fig.~\ref{fig:Ad_dynamU1}(b), we show the population evolution for slow tilting with $\alpha=0.001$. The dynamics is in the adiabatic limit, as $\sqrt{\alpha}\approx 0.03 \ll \Delta E$, leading to smooth population changes among the three wells. The system follows the ground state adiabatically, where the population tunnels from the leftmost to rightmost well.

For sufficiently strong nonlinear interactions, the lower levels develops loop structures near $U>U_{\rm{c}1}$. Due to the nonlinearity, the number of eigenvalues available is now greater than the dimension of the Hilbert Space. Dynamically, the system undergoes multiple avoided and direct level-crossings, when increasing $\gamma$ from $-\infty\to +\infty$. As a result, oscillations are seen in the diabatic regime due to the excitation of higher energy eigenstates [see Fig. \ref{fig:Ad_dynamU1}(c)]. In the adiabatic limit, the loop structures play vital roles as many eigenstates are excited, giving rise to extremely fast oscillations with multiple frequencies, as seen in Fig.~\ref{fig:Ad_dynamU1}(d).

To show the influence of the long-range interaction on the dynamics, we show the probability of the population being retained in the initial state (left well) for different tilting rate $\alpha$ in Fig.~\ref{fig:LZProb}. Interestingly, the excitation probability is largely captured by the Landau-Zener transition probability Eq.~(\ref{eq:LZ}) when the nonlinear interaction vanishes. When $U=1$ the retention probability of the initial well increases. It approaches to the non-interacting case in the adiabatic limit when $\alpha\ll 1$. Note that there are no simple power laws present in the tunneling probability as a function of $\alpha$, which is different from the double-well potentials~\cite{Liu2002a}.
For even stronger interactions $U=3$, the excitation probability depends on $\alpha$ non-monotonically. The retention probability is large for certain values of $\alpha$ when $\alpha<1$, and becomes dramatically larger when $\alpha>1$. As shown in Fig.~\ref{fig:eig}, the presence of the loop and level-crossings breaks the adiabatic condition. Violating the Landau-Zener prediction in the adiabatic limit has also been shown for both double-well~\cite{Liu2003a,Liu2002a,Albiez2005,Zibold2010} and triple-well system with short-range interactions~\cite{Guo2014,Graefe2006a}.

\subsection{Self-trapping and chaotic dynamics}\label{subsec:ST}
The retention probability in Fig.~\ref{fig:LZProb} indicates the emergence of self-trapping when the long-range interaction is strong. Self-trapping has been extensively studied in double-well potentials~\cite{Milburn1997, Liu2002a}. BECs with short-range interactions can localize in a single well as the densities scale logarithmically with the interaction strength, after a certain critical value. Self-trapping is also studied with short-range interacting BECs in triple-well potentials~\cite{Graefe2006a,Liu2007}. Here, we will discuss the differences between both short and long-range interactions, and how we can control the final distribution of atoms, by manipulating the initial conditions.

\figtop{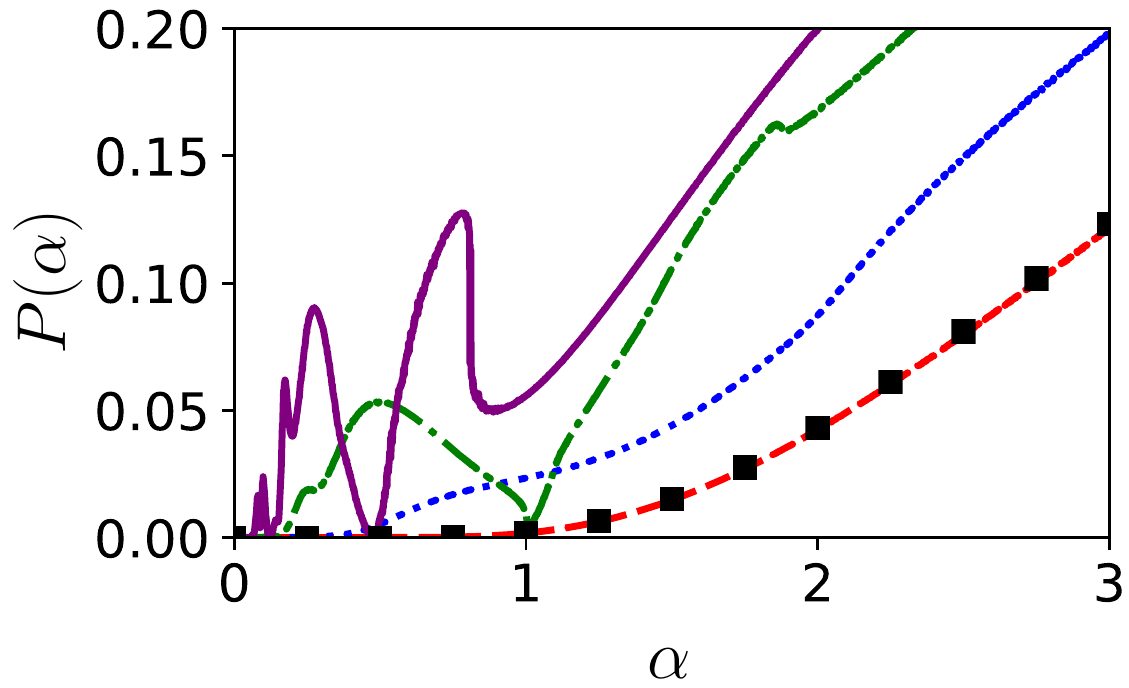}{(color online) \textbf{Retention probability of the initial state}. The analytical Landau-Zener probability for $U=0$ is given by the black squares. The other curves correspond to the numerical calculations for $U=0$ (red dashed), $U=1$ (blue dotted), $U=2$ (green dot-dashed), and $U=3$ (purple solid). The parameter $\gamma$ is varied from $-10\rightarrow+10$ in calculating the probability. }{0.62}{fig:LZProb}

Without nonlinear interactions ($U=0$), the mean-field Eqs. (\ref{eqm1}-\ref{eqm3}) are analytically solvable, yielding solutions
\begin{subequations}
\BE
c_1(t) \1 \frac{1}{\omega^2} \biggl\{  J\left[ J \bar{c}_1+ \gamma \bar{c}_2-J \bar{c}_3 \right]  \nn \\
\2 \left[ \gamma^2 \bar{c}_1 -J\gamma \bar{c}_2+J^2\left[\bar{c}_1+\bar{c}_3 \right] \right]\cos\omega t \nn\\
& + & \ii \omega\left[  J \bar{c}_2-\gamma \bar{c}_1 \right]
 \sin\omega t \biggl\},  \\
c_2(t) \1 \frac{1}{\omega^2}  \biggl\{  \gamma\left[ J \bar{c}_1+ \gamma \bar{c}_2-J \bar{c}_3 \right] \nn \\
\2 \left[ 2J \bar{c}_2 +\gamma[ \bar{c}_3-\bar{c}_1]  \right]\cos\omega t \nn\\
\2 \ii \omega[  \bar{c}_1- \bar{c}_3]  \sin\omega t  \biggl\},\\
c_3(t) \1 \frac{1}{\omega^2} \biggl\{ - J\left[ J \bar{c}_1+ \gamma \bar{c}_2-J \bar{c}_3 \right]  \nn\\
\2 \left[ J \gamma \bar{c}_2-\gamma^2 \bar{c}_3+J^2\left[ \bar{c}_1 +\bar{c}_3 \right] \cos\omega t  \right] \nn\\
 &+&   + \ii \omega \left[  J \bar{c}_2+\gamma \bar{c}_3 \right]
 \sin\omega t  \biggl\},
\EE\label{eq:ST}
\end{subequations}
where $\omega=\sqrt{2J^2+\gamma^2}$ and $Y^{\intercal}=[\bar{c}_1,\bar{c}_2,\bar{c}_3]$ denotes the eigenvectors at $t=0$.

In the presence of nonlinear interactions, the mean-field equations are solved numerically with a given set of parameters and initial conditions. To characterize dynamics in the long time limit, we calculate the time averaged densities in individual wells through
\BE
\braket{n_i}=\lim_{\tau\rightarrow\infty} \left[\frac{\int_0^{\tau}|c_i(t)|^2\dif t}{\tau}\right],
\EE
where $\tau$ is the final time. In the numerical simulations, we integrate the nonlinear GP equations from $t=0$ up to $\tau = 100$. We have checked that consistent results can be obtained when integrating the GP equations up to this time.

In the following we consider several different cases for both the symmetric and antisymmetric trap setups, to demonstrate the importance that the initial conditions have on the dynamics.
We begin by looking at the symmetric case where $\gamma=0$.

\subsubsection*{Case \RNum{1}: $Y^{\intercal}=[1/2,1/\sqrt{2},1/2]$}
The lowest energy eigenstate when $U=0$ is given by $Y^{\intercal}=[1/2,1/\sqrt{2},1/2]$. Using this as the initial state, the corresponding atomic densities in each wells are obtained by using Eq.~(\ref{eq:ST}),
\BE
\braket{n_1}=\braket{n_3}=\frac{1}{4} \text{, }\braket{n_2}=\frac{1}{2}.\nn
\EE
The majority of the particles are found in the middle well. Using the same initial state, we numerically solve the GP equations for different $U$. In Fig. \ref{fig:ST}(a), the average density decreases in the two outer wells while increasing in the middle site, as $U$ increases. When $U\gg1 $, the population tends to fully localize in the middle site. Due to strong nearest-neighbor and next-nearest-neighbor interactions, the lowest energy corresponds to all atoms sitting in one well, as we show in the numerical simulation.
Here we see a smooth transition from the initial densities towards the self-trapping regime. From Fig \ref{fig:eig_vsU} (a) we see that when $U>U_{c1}$ the lowest energy level is largely independent of $U$. The next excited level has also merged with the lowest level, preventing any occupation of higher energy modes. This accounts for the smooth increase in the densities as each the energy gap separating any higher levels is larger than the hopping strength, i.e $\Delta E >J$.

\subsubsection*{Case \RNum{2}: $Y^{\intercal}=[1,0,0]$ }
\begin{figure*}[t!]
\centering
\includegraphics[scale=0.65]{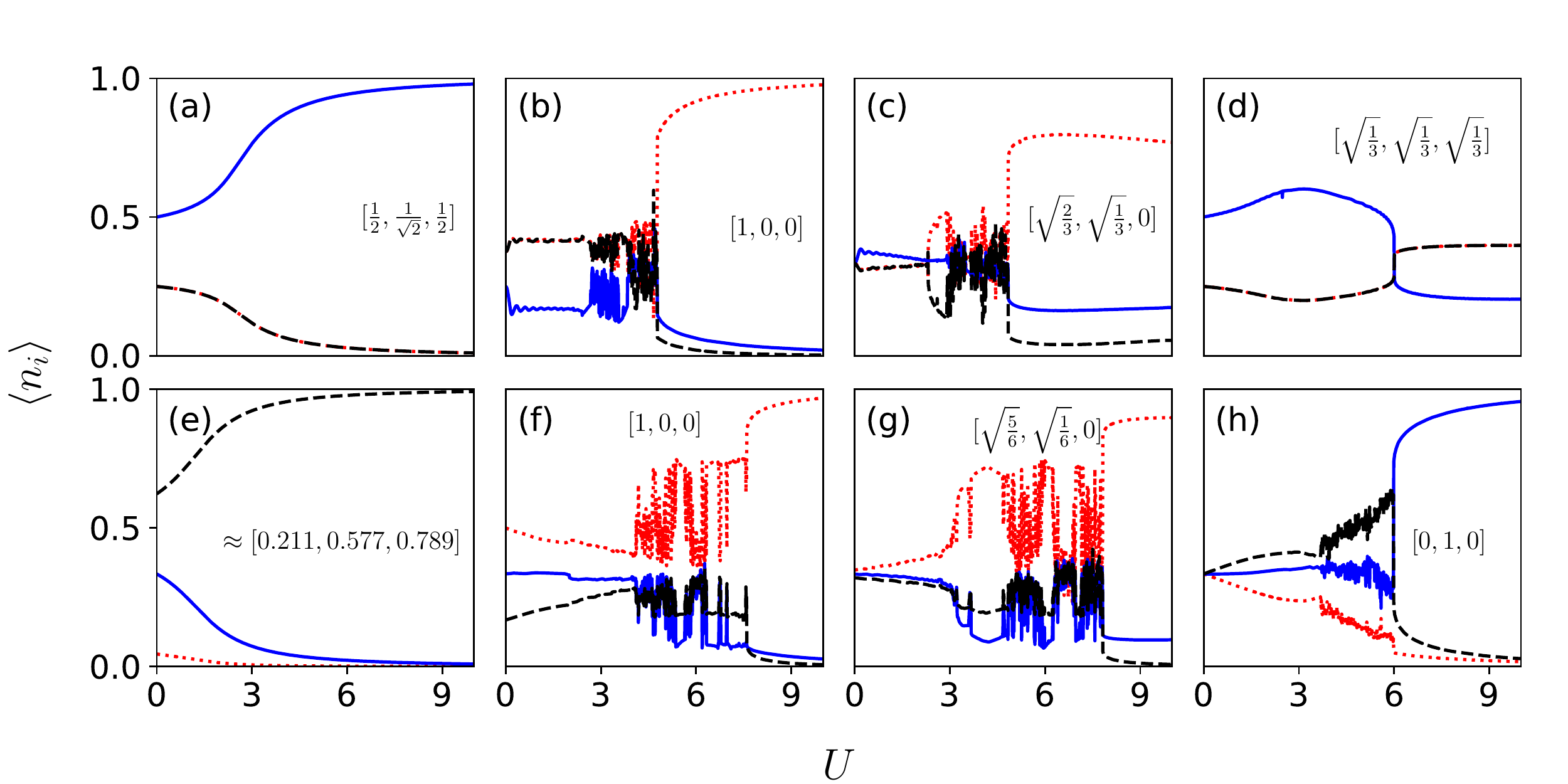}
\caption{(color online) \textbf{Self-trapping of the populations with different initial conditions}. The time-averaged densities of each site are shown as a function of the interaction strength $U$. The symmetric cases with $\gamma=0$ are shown in (a)-(d) and the tilted cases with $\gamma=1$ are shown in (e)-(h). The initial conditions $Y^{\intercal}=[\bar c_1,\bar c_2,\bar c_3]$ are shown as insets in each panel. The densities of the left, middle and right wells are denoted with red dotted, blue solid, and black dashed curves. The average density is obtained by evolving the GP equations to time $\tau=100$.}\label{fig:ST}
\end{figure*}

\noindent When changing the initial state to $Y^{\intercal}=[1,0,0]$, the dynamics of the population changes drastically. Without interactions, the average populations are obtained again with the help of Eq.~(\ref{eq:ST}),
\BE
\braket{n_1}=\braket{n_3}=\frac{3}{8} \text{, } \braket{n_2}=\frac{1}{4}\nn.
\EE
Increasing $U$, the average densities of the middle well decreases slightly and then stays at a lower value [Fig. \ref{fig:ST}(b)]. The populations then become turbulent as the interaction strength passes $U=U_{\rm{c}1}$, where the dynamics can not be categorized by standard Josephson or self-trapping regimes. Due to the complicated energy levels [see Fig.~\ref{fig:eig_vsU}(a)], chaotic dynamics is produced as particles tunnel between each site within the range of  $U_{c1}<U<U_{c2}$. This chaotic dynamics continues until the interaction strength passes $U=U_{\rm{c}2}$. The self-trapping re-emerges such that the BECs localize in the left well when $U>U_{c2}$.

\subsubsection*{Case \RNum{3}: $Y^{\intercal}=[\sqrt{2/3},\sqrt{1/3},0]$}
\noindent Fig \ref{fig:ST}(c) shows the system being initialized in the state $Y^{\intercal}=[\sqrt{2/3},\sqrt{1/3},0]$. Without interactions (i.e. $U=0$), densities in each well are
\BE
\braket{n_1}=\braket{n_2}=\braket{n_3}=\frac{1}{3}\nn.
\EE
As with the previous case, the dynamics is turbulent within the region of $U_{c1}<U<U_{c2}$ due to the superposition of energy levels. What is interesting is that the densities are no longer localized in a single well in the limit when $U\rightarrow\infty$. Here the weighting of the initial conditions have allowed for approximately $17\,\%$ of the atoms to occupy the middle well, with the remainder almost all in the left well.

\subsubsection*{Case \RNum{4}: $Y^{\intercal}=[\sqrt{1/3},\sqrt{1/3},\sqrt{1/3}]$}
\noindent Next we examine the initial condition  $Y^{\intercal}=[\sqrt{1/3},\sqrt{1/3},\sqrt{1/3}]$. The average density with $U=0$ is
\BE
\braket{n_1}=\braket{n_3}=\frac{1}{4}\text{, }\braket{n_2}=\frac{1}{2}\nn.
\EE
For this case we see drastically different dynamics in Fig. \ref{fig:ST}(d). First, we note that in the intermediate region  $U_{c1}<U<U_{c2}$, the system bypasses any chaotic dynamics. 
This will be examined in more detail in the coming section, however we can attribute this to the structure of phase space that the fixed points travel through.
Moreover, this case provides an example of self-trapping in three wells simultaneously, as $n_i >0 \text{ } \forall \text{ } i$, when $U\gg 1$.

\subsubsection*{Case \RNum{5}: $Y^{\intercal}\approx[0.221,0.577,0.789]$}
\noindent We now move on to examine the antisymmetric case by focusing on  $\gamma=1$. We begin by examining the lowest energy eigenvector $Y^{\intercal}\approx[0.221,0.577,0.789]$. When $U=0$ the average densities of each well are
\BE
\braket{n_1}\approx0.045 \text{, } \braket{n_2}=\frac{1}{3} \text{, }\braket{n_3}\approx 0.622\nn.
\EE
As with the symmetric case, the system is prepared in an eigenstate of the initial Hamiltonian, meaning that there is a smooth transition as the state follows the constant energy past $U\gtrsim3$. From Fig. \ref{fig:eig_vsU}(b) we see that the energy difference between the lowest energy state and the upper states is far larger than the hopping strength, i.e $\Delta E > J$, preventing coupling to higher energy states.
\subsubsection*{Case \RNum{6}: $Y^{\intercal}=[1,0,0]$}
\noindent We begin to see more interesting dynamics when the initial condition $Y^{\intercal}=[1,0,0]$ is again chosen. For the tilted wells this gives the noninteracting densities
\BE
\braket{n_1} =\frac{1}{2}\text{, } \braket{n_2} = \frac{1}{3} \text{, }\braket{n_3} =  \frac{1}{6}\nn.
\EE
At first glance these initial values may seem uninteresting, however they imply that even though the trap is orientated such that the rightmost well has the lowest overall level bias, the densities are still localized mainly in the leftmost well. This phenomena is extremely counter-intuitive as one would expect a large proportion of the densities to tunnel to the lowest available state. When we numerically solve the nonlinear GP equation [see Fig. \ref{fig:ST}(f)], we see this feature persist for strong nonlinear interaction strength $U>\overline{U}_{c3}$. The intermediate chaotic region now spans the entire range of $\overline U_{c1}<U<\overline U_{c3}$, as the tilted orientation produce a further energy level at much larger interaction strengths [see Fig. \ref{fig:eig_vsU}(b)]. As $U\rightarrow\infty$, we see that the localization is almost fully in the leftmost well, with the highest level bias energy.
Similar phenomena where reported for the short-range interacting system in Ref. \cite{Liu2007}.
\subsubsection*{Case \RNum{7}: $Y^{\intercal}=[\sqrt{5/6},\sqrt{1/6},0]$}
\noindent The noninteracting density for this case can be obtained by using Eq~(\ref{eq:ST}),
\BE
\braket{n_1} = \frac{17-2\sqrt{5}}{36}\text{, } \braket{n_2} = \frac{1}{3} \text{, }\braket{n_3} = \frac{17+2\sqrt{5}}{36}\nn.
\EE
Similar to the previous case in Fig. \ref{fig:ST}(g), we see that this initial condition yields highly chaotic dynamics, where the range of the chaos extends the region $\overline U_{c1}<U<\overline U_{c3}$.
When the nonlinear interaction is strong, and the system enters the self-trapped regime ($U>\overline U_{c3}$) we see that self trapping occurs in the leftmost and middle wells, with roughly 10\% of the particles occupying the middle site.

\subsubsection*{Case \RNum{8}: $Y^{\intercal}=[0,1,0]$}
\noindent In this case, the density when $U=0$ is
\BE
\braket{n_1}=\braket{n_2}=\braket{n_3}=\frac{1}{3}\nn.
\EE
\noindent In Fig. \ref{fig:ST}(h) we see self-trapping dynamics once the interaction passes $U>\overline U_{c2}$, as only the lowest energy level is occupied. When $\overline U_{c1}<U<\overline U_{c2}$ the dynamics is unstable such that the average density fluctuates drastically when varying $U$.

\figtop{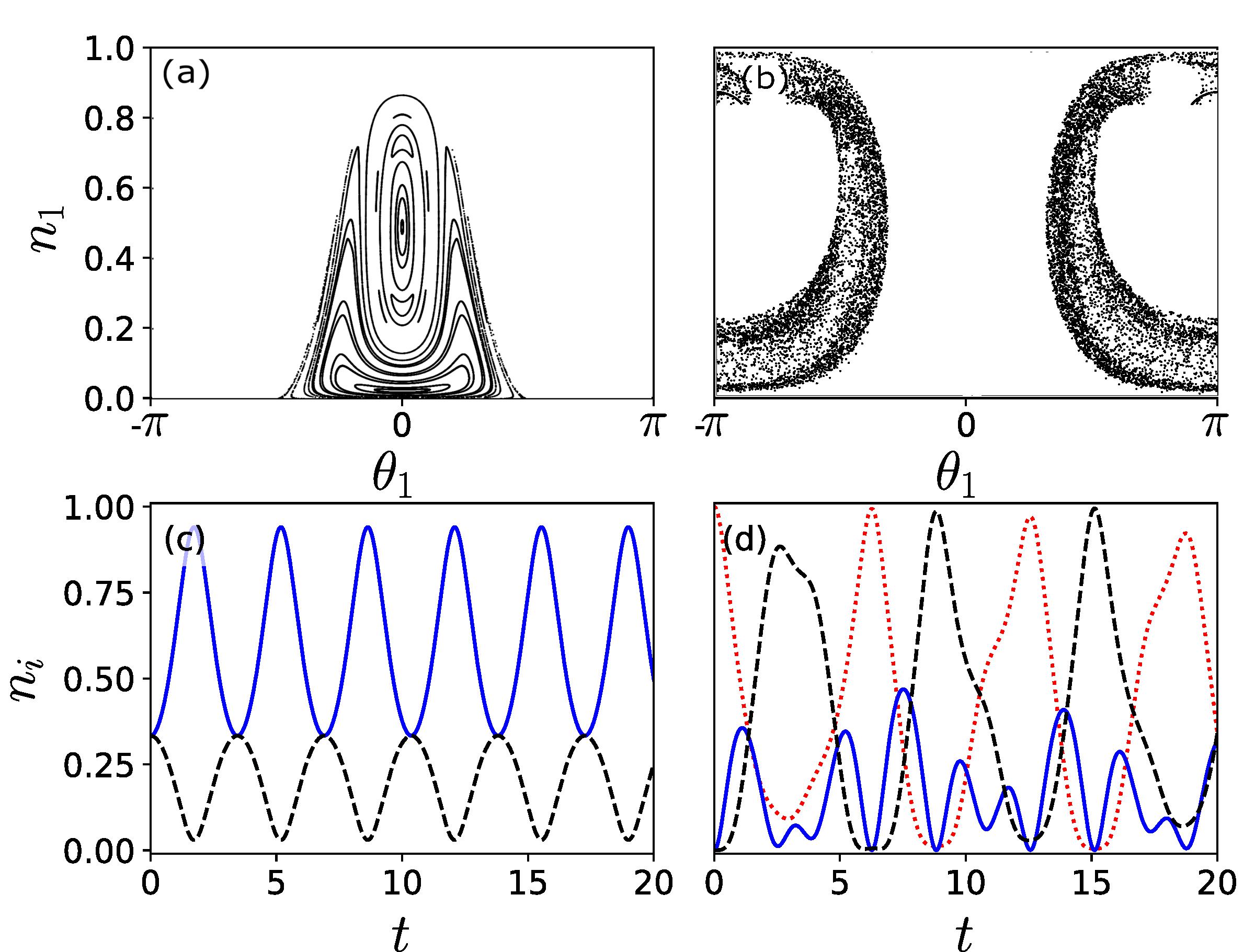}{(color online) \textbf{Poincar\'e Sections, regular and chaotic population dynamics}. The Poincar\'e sections for (a) $E=-0.5$ and (b) $E=0.2$ are shown. Panels (c) and (d) show dynamics of $n_1$ (red dotted), $n_2$ (solid blue), and $n_3$ (black dashed) using initial conditions that would lie on the sections of (a) and (b) respectively. Other parameters are $U=3$ and $\gamma=0$.}{0.42}{fig:PS}
\subsection{Poincar\'e Sections and chaotic dynamics}
Fig.~\ref{fig:ST} shows that regular and chaotic dynamics can be triggered by varying the initial state, even when the long-range interactions are the same. This dependence comes from the fact that energies of the system are changed when considering different initial states. As the energy is a conserved quantity, the system will show vastly different trajectories in phase space. We now illustrate this dependence using Poincar\'e sections \cite{Liu2007,Franzosi2003}.

To obtain the Poincar\'e section, Hamiltonian (\ref{Ham:mf}) is converted to a classical Josephson-like analogue, where the resulting equations of motion define a 4-dimensional phase space in terms of the canonical variables $\{n_1,\theta_1\}$ and $\{n_3,\theta_3\}$ [see appendix \ref{appendix:PS} for more details]. By taking a slice at $\theta_3=0$, in the direction of $\dot{\theta_3}<0$, and employing energy conversation, the equations of motion can be expressed inside the 2-dimensional plane $\{n_1,\theta_1\}$, forming the Poincar\'e section.

In Fig. \ref{fig:PS}(a) we show the Poincar\'e section when the average energy $E=\langle \tilde{H}\rangle/N=-0.5$. Regular orbits mean that solutions to the dynamics will travel across phase space via smooth paths periodically. This energy is associated with the initial conditions given by Figs. \ref{fig:ST}(a) and (d), which do not show chaos in their time-averaged dynamics in the interplay region of $U_{c1}<U<U_{c2}$. In Fig. \ref{fig:PS}(c) we show dynamics of the population that corresponds to the initial condition of Fig. \ref{fig:ST}(d). The periodic oscillation of the population is consistent with the regular pattern in the Poincar\'e section. Fig. \ref{fig:PS}(b) shows a very different situation where the Poincar\'e section at $E=0.2$ only has localized regions of chaos, corresponding to the initial conditions of Figs. \ref{fig:ST} (b) and (c). In Fig. \ref{fig:PS}(d), we see that the associated dynamics does not shown regular periodic oscillations. Recent studies have found interesting chaotic dynamics emerging from three-state systems when nonlinear interactions become strong~\cite{Dey2018,Dey2019,Burkle2019}. The understanding of the chaotic dynamics and its control in Rydberg-dressed BECs will be useful for future experiments.

\subsection{Comparison between quantum and mean-field dynamics}
The mean-field dynamics presented in previous sections is obtained in the limit $N\to \infty$. Experimentally, self-trapping of populations has been observed with BECs containing about $1000$ atoms in double-well potentials, where dynamics of the BEC can be accurately described by the mean-field theory~\cite{Albiez2005}. In this section, we will show that the adiabatic and nonadiabatic dynamics predicted by the mean-field theory can be also seen in relatively small systems with $N\le 100$. To study the quantum dynamics, we numerically solve the  Schr\"odinger equations using the three-site Bose-Hubbard Hamiltonian~(\ref{Ham:bh}). We will encounter a time-dependent Hamiltonian when studying the Landau-Zener transition. 

\figtop{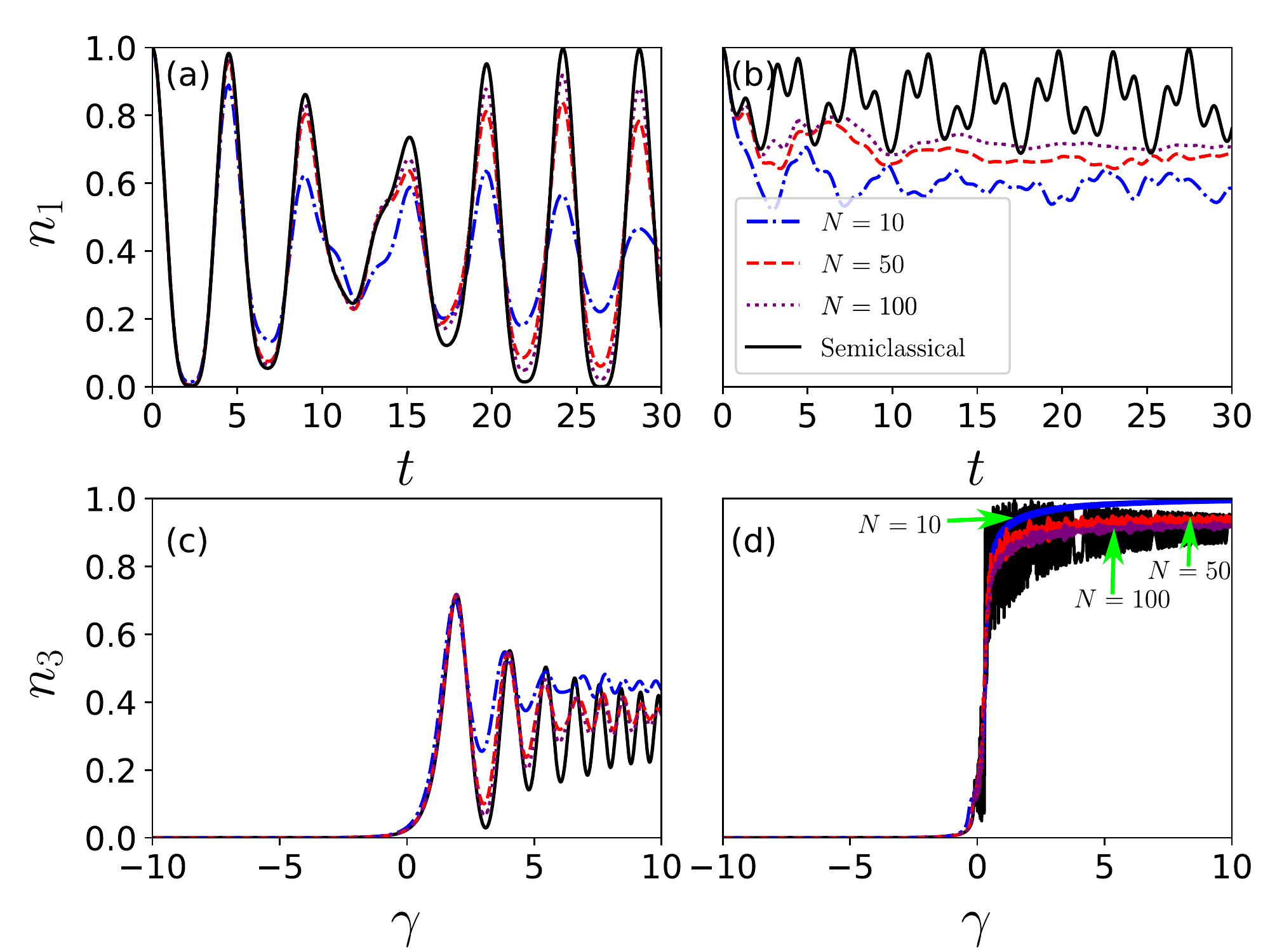}{(color online) \textbf{Quantum and semiclassical dynamics}. Populations obtained from the extended Bose-Hubbard Hamiltonian~(\ref{Ham:bh}) using (a) $U=1$ and (b) $U=5$ for different atom numbers. The black solid curves show the mean-field results with same interactions.  Landau-Zener transitions are shown when (c) $\alpha=1$ and (d) $\alpha=0.01$ with $U=3$. Arrows are used in (d) to distinguish the different oscillations in gray scale.}{0.42}{fig:quantum}

In the Josephson oscillation regime when $U<U_{c1}$, populations oscillate among the potential wells almost completely, as shown in Fig. \ref{fig:quantum}(a). At later times, the population partially returns to the initial well. The long-time dynamics of the population shows a relaxation, especially when $N$ is small. Increasing $N$, we find that the relaxation becomes weaker, such that the quantum dynamics resembles that of the mean-field calculation. Similar dynamics has been investigated in detail in Ref \cite{Paredes2009}. When approaching to the self-trapping regime ($U>U_{c1}$), only a small fraction of populations can tunnel to other potential wells. The population oscillates irregularly around a constant that is close to 1 [see Fig. \ref{fig:quantum}(b)]. Increasing $N$, we find that amplitudes of the oscillation decrease rapidly, and the average population also increases. The average population, however, is smaller than the mean-field result. The difference is largely attributed to the many-body correlations between potential wells, which are neglected in the mean-field calculations

In Fig. \ref{fig:quantum}(c) and (d) we study Landau-Zener dynamics by dynamically changing the trap bias from $\gamma=-10$ to $\gamma=10$ in Hamiltonian~(\ref{Ham:bh}). The corresponding mean-field dynamics is shown in Fig \ref{fig:Ad_dynamU1}(c) and (d). When rate $\alpha$ is large, the right well starts to be populated once $\gamma >0$. Further increasing $\gamma$, the population oscillates with larger amplitudes for larger $N$. Remarkably, such evolution agrees with the mean-field calculation well. In the adiabatic regime with $\alpha=0.01$, the mean-field calculation shows rapid oscillations around  $n_3\sim 1$. We note that the quantum dynamics is less oscillatory than the mean-field result, especially when $N$ is large. However, asymptotic values from both quantum and mean-field calculations agree when $\gamma\gg1$. 
\section{Conclusion}\label{sec:Conclusion}
We have studied the dynamics of Rydberg-dressed BECs in a triple well potential.
Within the mean-field theory, we have obtained eigenenergies of the system for different combinations of parameters. It is found that the eigenspectrum develops multiple level-crossings in the lower branches of the eigenspectra, when the soft-core interaction is strong. The presence of level-crossings in the lower branches leads to more complicated dynamics than BECs with only short-range interactions. We have shown that it is possible to achieve self-trapping of populations in either one, two, or three wells by varying the initial conditions and the level bias. We have identified parameter regions, where dynamics is chaotic. This is demonstrated with the population evolution, and further confirmed with Poincar\'e sections. By numerically solving the quantum Hamiltonian for fixed particle numbers, we have shown that the mean-field results can be largely observed when the particle number $N\sim 100$.
In the future, it is interesting to study how chaos emerges in the finite trap system due to strong long-range interactions. Moreover, it would be advantageous to increase the number of sites to explore mean-field and quantum mechanical effects due to the soft-core interaction. In large and tilted lattices, one could also study Bloch oscillations of BECs with strong and long-range interactions.

\section*{Acknowledgements}
We thank Dominic Rose and Filippo Gambetta for fruitful discussions. The research leading to these results received
funding from EPSRC Grant No. EP/R04340X/1 via the
QuantERA project “ERyQSenS,” UKIERI-UGC Thematic
Partnership No. IND/CONT/G/16-17/73, and the Royal Society
through International Exchanges Cost Share Award No.
IEC$\backslash$NSFC$\backslash$181078. We are grateful for access to the Augusta
High Performance Computing Facility at the University of
Nottingham.

\begin{appendix}

\renewcommand{\thefigure}{A\arabic{figure}}
\setcounter{figure}{0}

\begin{figure}[t!]
\centering
\includegraphics[scale=0.42]{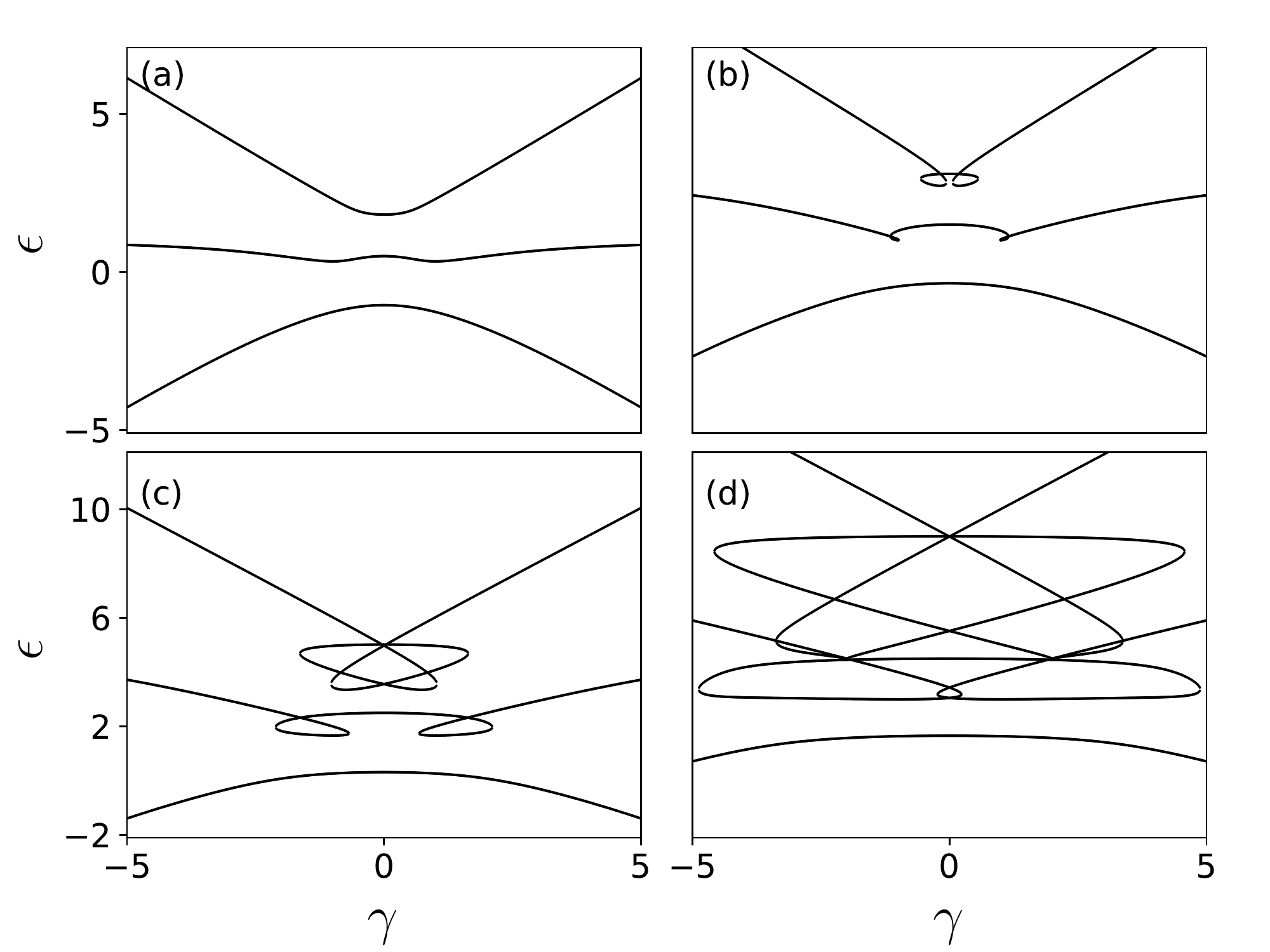}
\caption{\textbf{Eigenspectra with barely short-range interactions}. We show eigenspectra for nonlinear systems with only short-range interactions for (a) $W=1$, (b) $W=3$, (c) $W=5$, and (d) $W=9$ while fixing $U=V=0$. The level-crossings are only found in the upper branches.}
\label{figA1}
\end{figure}

\section{Symmetry between short-range and long-range interacting systems}\label{appendix:SR_vs_LR}
\begin{figure}
\centering
\includegraphics[scale=0.42]{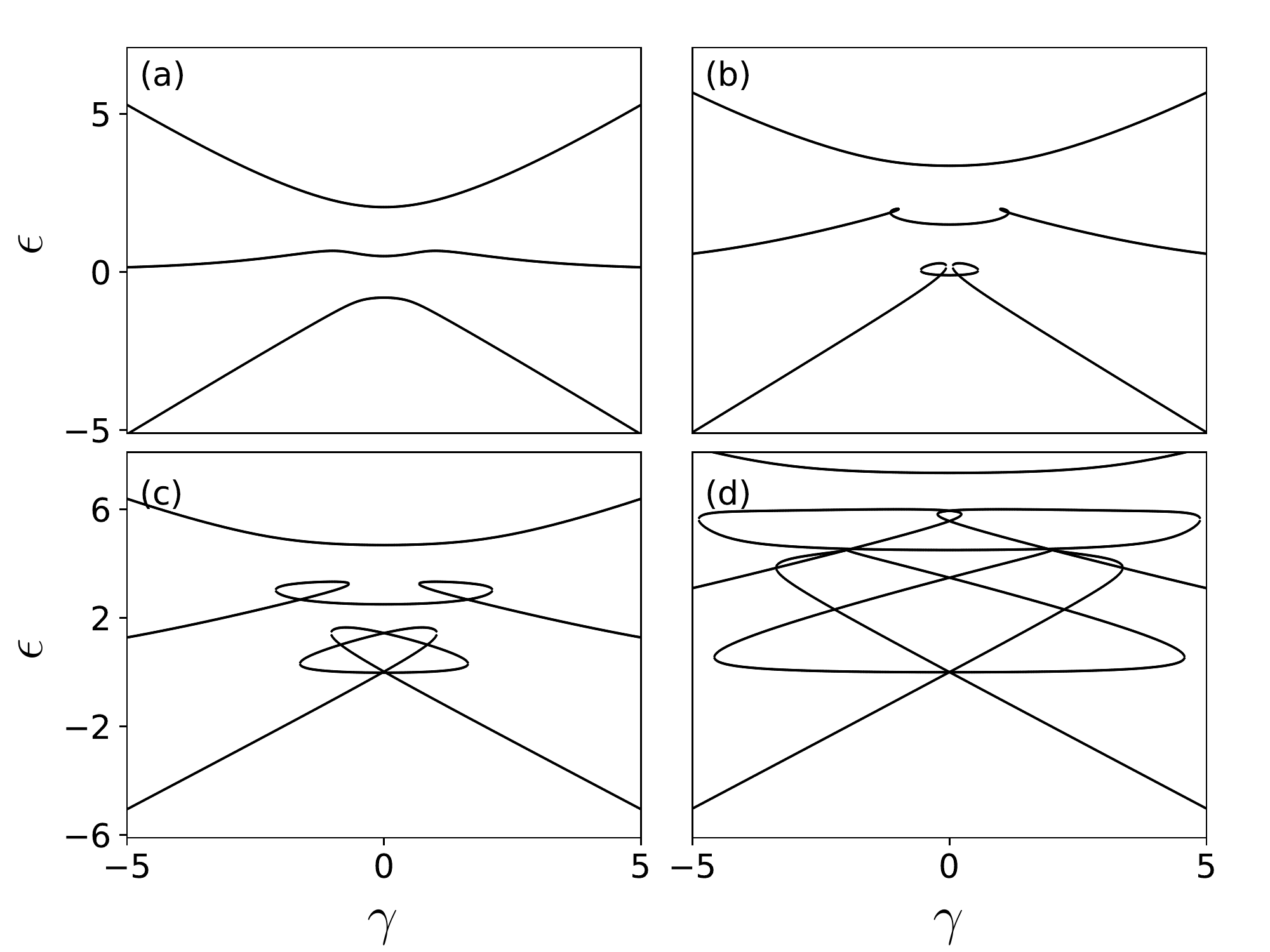}
\caption{\textbf{Eigenspectra when $U=V$}. The eigenspectra are shown for (a) $U=V=1$, (b) $U=V=3$, (d) $U=V=5$ and (d) $U=V=9$.}
\label{figA2}
\end{figure}

If only the short-range interaction is present in the system, the corresponding nonlinear GP equations read
\begin{subequations}
\BE
i \dot{c}_1 \1  - Jc_2  + W|c_1|^2c_1+\gamma  c_1,\label{app_eqm1}\\
i \dot{c}_2 \1 -J(c_1+c_3) + W|c_2|^2 c_2,\label{app_eqm2}\\
i \dot{c}_3 \1  - Jc_2 + W|c_3|^2c_3 -\gamma c_3.\label{app_eqm3}
\EE
\end{subequations}
The respective eigenspectrum shows complicated level-crossings when the onsite interaction $W$ is strong, as shown in Fig.~\ref{figA1}. Note that these structures only show in the upper branches.

Now consider a special situation with $V=U$ and $W=0$, the GP equations become,
\begin{subequations}
\BE
i \dot{c}_1 \1  - Jc_2  - U|c_1|^2c_1+(U+\gamma)  c_1,\label{app_eqm4}\\
i \dot{c}_2 \1 -J(c_1+c_3) - U|c_2|^2c_2+Uc_2,\label{app_eqm5}\\
i \dot{c}_3 \1  - Jc_2 - U|c_3|^2c_3 +(U-\gamma) c_3,\label{app_eqm6}
\EE
\end{subequations}
where we have used the normalization condition $\sum_{j}|c_j|^2=1$.
This means that long-range repulsive interactions are equivalent to short-range attractive interactions (plus a global energy shift $U$). The symmetry of the system will not be changed when we change the sign of parameter $J$. Hence the eigenspectra of the long-range interacting BEC can be obtained by flipping Fig.~\ref{figA1} (after shifting downwards by $U$). This can be seen from our numerical calculation, shown in Fig. \ref{figA2}. This also explains why the level-crossings emerge in the lower branches in the main text.

\renewcommand{\thefigure}{B\arabic{figure}}
\setcounter{figure}{0}

\section{Landau-Zener dynamics with short-range interactions}\label{appendix:SR}

At first glance, the differences between Fig. \ref{figA1}(c) and Fig. \ref{figA2}(c) may not be apparent. However, the fact that the bifurcation of the energy levels happens on the lowest energy state for the long-range interacting system leads to dramatically different physics when compared to its short-range counterpart.
\begin{figure}[h!]
\centering
\includegraphics[scale=0.4]{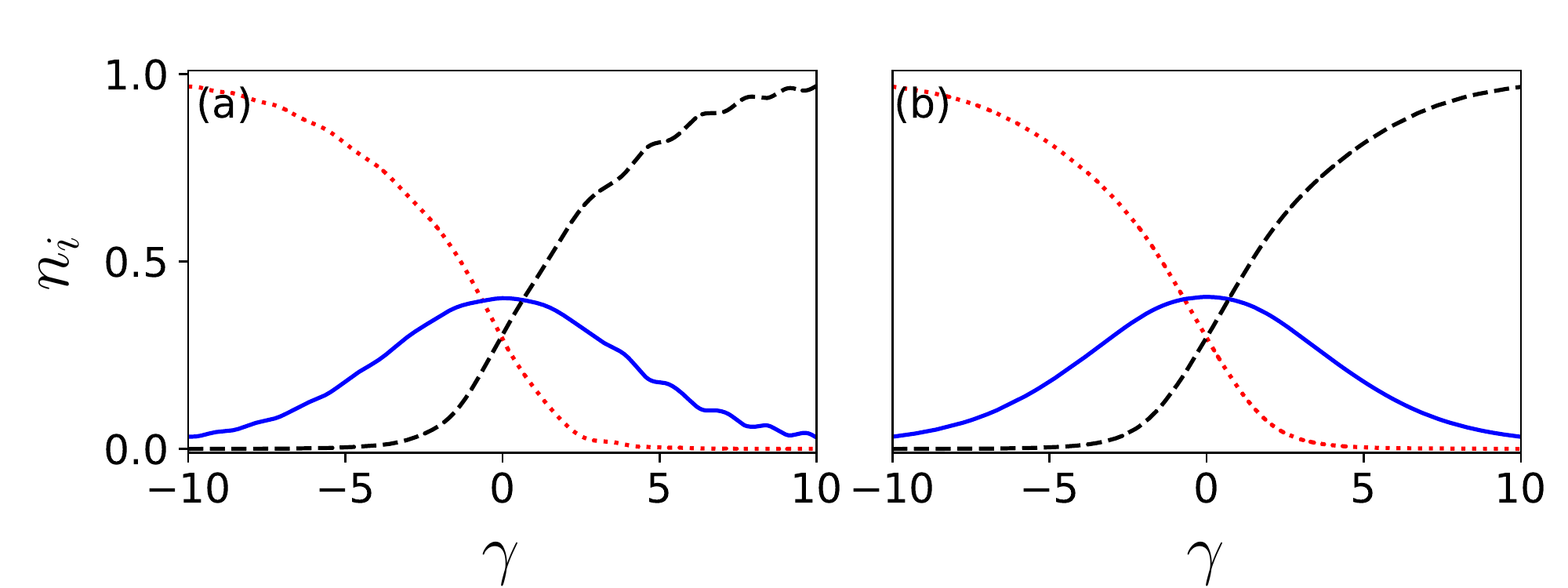}
\caption{(color online) \textbf{Landau-Zener Dynamics for a short-range interacting system.} The interaction strength $U=5$ for both panels. Here (a) is $\alpha=1$ and (b) $\alpha=0.001$. The system is initialize in the left well. }
\label{figA3}
\end{figure}
 In the short-range case, the system is allowed to follow a complete adiabatic transfer of the populations as there is no splitting of the ground state eigenspectrum as the tilt moves from $\gamma =-\infty$ to $\gamma =+\infty$. This can be seen explicitly when we evolve the time dependent nonlinear GP equation for the short-range system, and calculate the Landau-Zener dynamics, as we did in Sec. \ref{subsec:LZ} of the main text.

From Fig. \ref{figA3}(a) we see that the fast quench produces minor oscillations in the densities, but even at this speed there is almost a complete transfer from the leftmost well to the rightmost well. In Fig. \ref{figA3}(b), where we examine the slow quench, the system complete responds to the changes in the eigenenergies and a complete transfer is seen. This shows the short-range interactions produce quantitatively different physics compared to long-range interactions. 

\renewcommand{\thefigure}{C\arabic{figure}}
\setcounter{figure}{0}

\section{Canonical Representation of Phase Space}\label{appendix:PS}
The amplitudes of each site can be expressed in terms of the total density and a phase factor as $c_i=\sqrt{n_i}\ee{i\phi_i}$. Importantly, only the relative phase between each site is an observable, meaning we can define the relative phase factors $\theta_1=\phi_2-\phi_1$ and $\theta_3=\phi_2-\phi_3$. The conservation condition means that the densities of the second site is defined by $n_2=1-n_1-n_3$.
Using these, the mean-field Hamiltonian [Eq. (\ref{Ham:mf}) of main text] can be expressed similarly to a classical Josephson Hamiltonian of the form
\begin{widetext}
\BE
\mathcal{H} =  - 2J\sqrt{1-n_1-n_3}\left[\cos(\theta1)\sqrt{n1}+\cos(\theta3)\sqrt{n3}\right]+ U(1-n_1-n_3)(n_1+n_3) +Vn_1n_3 +\gamma(n_1-n_3)
\EE
The resulting Lagrangian equations of motion for conserved momenta then read
\BE
\dot n_1 \1 - 2J\sqrt{n_1}\sqrt{1-n_1-n_3}\sin(\theta_1)\\
\dot n_3 \1 - 2J\sqrt{n_3}\sqrt{1-n_1-n_3}\sin(\theta_3)\\
\dot \theta_1 \1 U\left(1-2n_1-2n_3 \right) +V n_3 +\gamma - \frac{J\sqrt{1-n_1-n_3}\cos(\theta_1)}{\sqrt{n_1}} +\frac{J\left[ \sqrt{n_1}\cos(\theta_1)+\sqrt{n_3}\cos(\theta_3)\right]}{\sqrt{1-n_1-n_3}}\\
\dot \theta_3 \1 U\left(1-2n_1-2n_3 \right) +V n_1  - \gamma - \frac{J\sqrt{1-n_1-n_3}\cos(\theta_3)}{\sqrt{n_3}} +\frac{J\left[ \sqrt{n_1}\cos(\theta_1)+\sqrt{n_3}\cos(\theta_3)\right]}{\sqrt{1-n_1-n_3}}
\EE
\end{widetext}

These equations provide an alternate way of calculating the dynamics, which can also be used to explore how the relative phase of each site changes as a function of time. For the purposes of this work, we use these equations to calculate the Poincar\'e sections. For a given set of initial condition $\{ n_1(0),\theta_1(0)\}$, conversation of energy allows us to find the initial $n_3(0)$ for a given energy value $E$ where $E=\mathcal{H}$, while looking along the plane of $\theta_3=0$. The intersection of $n_1$ and $\theta_1$ along the plane of $\theta_3=0$ in the $\dot{\theta}_3<0$ are recorded to produce the Poincar\'e section.

\end{appendix}


%

\end{document}